\documentclass[aps,twocolumn,floats,prd,nofootinbib,superscriptaddress,longbibliography]{revtex4-1}
\usepackage[utf8]{inputenc}
\usepackage{graphicx,amsmath,amsfonts,amssymb,slashed,bbm,euscript,capt-of}
\usepackage{bbold,wasysym}
\usepackage{xcolor}
\usepackage{color}
\usepackage{hyperref}
\hypersetup{colorlinks, citecolor=blue, linkcolor=red, urlcolor=blue}
\usepackage{lipsum}
\usepackage[english]{babel}
\usepackage{blindtext}
\newcommand{\bs}{\boldsymbol}
\def\apjl{\rm ApJ}				
\DeclareUnicodeCharacter{2212}{-}	

\begin{document}

\title{Probing Gravitational Slip with Strongly Lensed Fast Radio Bursts}

\author{Tal Adi}
\affiliation{Department of Physics, Ben-Gurion University of the Negev, Be'er Sheva 84105, Israel}

\author{Ely D. Kovetz}
\affiliation{Department of Physics, Ben-Gurion University of the Negev, Be'er Sheva 84105, Israel}

\begin{abstract} 

The rapid accumulation of observed Fast Radio Bursts (FRBs) originating from cosmological distances makes it likely that some will be strongly lensed by intervening matter along the line of sight. Detection of lensed FRB {\it repeaters}, which account for a noteworthy fraction of the total population, will allow not only an accurate measurement of the lensing time delay, but also follow-up high-resolution observations to pinpoint the location of the lensed images. Recent works proposed to use such strongly-lensed FRBs to derive constraints on the current expansion rate $ H_{0} $ as well as on cosmic curvature. Here we study the prospects for placing constraints on departures from general relativity via such systems. Using an ensemble of simulated events, we focus on the gravitational slip parameter $\gamma_{\rm PN}$ in screened modified gravity models and show that FRB time-delay measurements can yield  constraints as tight as $ \left| \gamma_{\rm PN}-1\right| \lesssim 0.04\times(\Lambda/100\rm kpc)\times[N/10]^{-1/2} $ at $1\sigma$ with $10$  detections. 
\end{abstract}

\maketitle

\section{Introduction}

The accelerated cosmic expansion is one of the most puzzling open questions in cosmology and physics in general \cite{SupernovaSearchTeam:1998fmf,SupernovaCosmologyProject:1998vns}. The simplest solution is given by the introduction of a cosmological constant into Einstein's General Relativity (GR) equations, which leads to an exponential expansion once the Universe expands enough for the energy density in radiation and then matter to become subdominant. However, the physical interpretation for this ad-hoc solution remains unknown. Numerous theories propose an explanation for the acceleration \cite{Weinberg:1988cp,Huterer:2017buf,Caldwell:2009ix,Joyce:2016vqv} and can be categorized into two general approaches: dark energy (DE) and modified-gravity (MG), which can be thought of as two different ways to alter Einstein's equations in order to accommodate a similar effect to that of the cosmological constant. The former introduces a DE component with a modeled equation of state, which describes its properties and dynamics, thus altering the source terms of the equations, whereas the latter offers to modify the coupling to gravity, resulting in an ``effective'' Einstein tensor. While DE models can be evaluated via their equation of state, MG models are usually evaluated via an extension of the post-Newtonian parameterization to cosmological scales. Unfortunately, there is no one-description-to-rule-them-all and one must describe the model in order to define its post-Newtonian parameters.

Nonetheless, most MG theories share two key features. One is the \emph{gravitational slip}, referring to the difference between the potential associated with the time component of the metric and the one associated with the spatial component \cite{Daniel:2008et,Bertschinger:2011kk}. The second  is \emph{screening}, which sets a scale beyond which MG is relevant~\cite{Vainshtein:1972sx,Babichev:2013usa}.
Recent work, Ref.~\cite{Jyoti:2019pez}, took advantage of these features in order to demonstrate how one may place constraints on MG models using time-delay measurements of strongly-lensed quasars by adopting a phenomenological approach in which the departure from GR is encapsulated by a parameterized gravitational slip on distances greater than a cutoff scale $\Lambda$. However, the constraints are limited by the $\gtrsim 1\%$ uncertainty in the measured quasar time delay. One way to overcome this is to use a different source.

Fast Radio Bursts (FRBs) are bright transients with $\mathcal{O}(1\,{\rm ms})$ duration at $\sim$GHz frequencies. While the physical nature of FRBs remains a mystery \cite{Lorimer:2007qn,Thornton:2013iua}, their all-sky distribution and their large dispersion measures (DMs) are consistent with a cosmological origin \cite{Thornton:2013iua}. Some of the detected bursts feature a repeating pattern, enabling high-time-resolution radio interferometric follow-up observations to localize their sources \cite{Chatterjee:2017dqg}. The CHIME/FRB Collaboration recently presented a catalog \cite{CHIMEFRB:2021srp} of 535 FRBs over a year of observation, 61 of which originate from 18 repeating sources. 
Detecting a repeating FRB signal for which each burst is accompanied by a second image separated by the same fixed $\mathcal{O}(10)$ days time interval could indicate a repeating strongly-lensed FRB event. Events like this, which can be picked by  a large-field survey, can then be subsequently observed at higher resolution using radio telescopes such as the Very Large Array (VLA)~\cite{Condon:1998iy}, the future Square Kilometre Array (SKA)~\cite{5136190} or with Very Long Baseline Interferometry (VLBI) networks, enabling to resolve the lensed images of the FRB signal.
The frequent all-sky rate of FRBs, $\sim10^{3}$ per day \cite{ANTARES:2017hvn,Petroff:2019tty}, and their extremely short relative durations ($\sim 10^{-9}$ of a typical lensing time delay) make strongly-lensed repeating FRBs a promising tool for cosmological tests~\cite{Munoz:2016tmg,Li:2014dea,Dai:2017twh}. 
To that effect, recent work \cite{Li:2017mek} simulated such possible strongly-lensed FRB in order to conduct cosmography, and estimated the constraining power of such systems on the Hubble parameter $H_{0}$ and the spatial curvature $\Omega_{k}$ in a cosmological-model-independent manner, finding that 10 such systems can constrain $ H_{0} $ to a sub-precent level, on par with the tight constraints from the cosmic microwave background (CMB) measurement reported by Planck 2018 \cite{Planck:2018vyg}.

In this work we study the prospects of using strongly-lensed FRB systems to place constraints on screened MG models featuring a gravitational slip. While such systems have not yet been detected, it is likely that such detection will be made in the near future as the search for FRB expands. For example, the SKA \cite{5136190}, the Hydrogen Intensity and Real-time Analysis eXperiment (HIRAX) \cite{Newburgh:2016mwi}, and the Packed Ultra-wideband Mapping Array (PUMA) \cite{PUMA:2019jwd}, are all future experiments with wide fields of view which are expected to detect $ \sim10^{2}-10^{3} $ FRBs per day. Considering a typical galaxy-galaxy strong-lensing optical depth of $\sim10^{-4}-10^{-3}$~\cite{Liu:2019jka}, this leads to a predicted number of $ \mathcal{O}(10-100)$ lensed FRB events per year. As a considerable fraction of FRBs are repeaters, we optimistically assume (as in Ref.~\cite{Li:2017mek}) that within a few years $ \mathcal{O}(10)$ strongly-lensed repeating FRB systems can be detected and followed-up with high-resolution images.

To estimate the constraining power of FRB TD measurements on the Post-Newtonian gravitational slip parameter $ \gamma_{\rm PN}$ at super-galactic screening scales $\Lambda$, we adopt a formalism which resembles the one in Ref.~\cite{Jyoti:2019pez}. Using simulations of strongly-lensed FRB systems, taking into account the possibility of gravitational slip, we find that their observation will improve the constraints on $ \gamma_{\rm PN} $ significantly. With $ N\!=\!10 $ systems, the constraint scales as $ \left| \gamma_{\rm PN}-1\right| \lesssim 0.04\times(\Lambda/100\rm kpc)\times[N/10]^{-1/2} $, enabling $ <10\% $ precision up to scales of $ \Lambda\sim 300\rm kpc $.

The paper is organized as follows. In Section~\ref{sec:methods} we review the model we use to evaluate the TD and the methods we use to simulate strongly-lensed FRB systems. We then present and discuss our results in Section~\ref{sec:results}, before concluding in Section~\ref{sec:conclusions}.

\section{Methodology}\label{sec:methods}

\subsection{The Model}\label{subsec:model}

The metric on cosmological scales, in the presence of a gravitational potential, is described by the line-element
\begin{equation}
ds^{2} = a^{2}(\eta) \left[ - (1 + 2\Phi)d\eta^{2} + (1 - 2\Psi)d\vec{x}^{2} \right], \label{eq:linelement}
\end{equation}
where $ \Phi $ and $ \Psi $ are the Newtonian and longitudinal gravitational potentials, and $ a(\eta) $ is the cosmological scale factor. This metric gives rise to the familiar Newton's equation, $ \vec{x}'' = -\vec{\nabla}^{2}\Phi $, and Poisson's equation, $ \nabla^{2}\Psi = 4\pi G a^{2} \rho $. While the two potentials are equal under GR \cite{1992grle.book.....S,Ma:1995ey}, MG theories, such as $ f(R) $ \cite{Sotiriou:2008rp,Capozziello:2003tk} and scalar-tensor theories \cite{2004sgig.book.....C,Nojiri:2010wj,Schimd:2004nq,Nojiri:2017ncd}, DGP gravity \cite{Dvali:2000hr,Lue:2005ya,Song:2006jk}, and massive gravity \cite{Dubovsky:2004sg}, all predict a systematic difference $ \Psi\ne\Phi $, also known as the gravitational slip. This departure from GR is often quantified by the ratio $ \gamma_{\text{PN}} = \Psi/\Phi $, while $ \gamma_{\text{PN}}=1 $ (i.e.\ GR) is expected at small distances due to screening. It is worth noting, though, that many efforts to develop a phenomenological description of the gravitational slip result in a variety of parameterizations, including dynamical ones~\cite{Hu:2007pj,Amendola:2007rr,Caldwell:2007cw,Zhang:2007nk,Daniel:2008et,Zhang:2008ba}.

The gravitational screening phenomenon reflects in the suppression of the additional gravitational degrees of freedom, introduced by MG theories, within a certain region. This effect enables MG theories to resemble GR at sub-galactic scales, while introducing new effects at larger scales, such as cosmic acceleration. We follow the notation in Ref.~\cite{Jyoti:2019pez}, and consider the gravitational slip to be turned on abruptly at a screening radius $ \Lambda $, which is assumed to be larger than the Einstein radius,  $ \Lambda > \theta_{E}D_{L} $.

Photons follow null geodesics, $ ds^{2}=0 $, which, according to Eq.~\eqref{eq:linelement}, are described by $ \Sigma \equiv \Phi + \Psi $. Thus, for a spherically-symmetric mass distribution, we can model the departure from GR as \cite{Jyoti:2019pez}
\begin{equation}
\Sigma = \left[ 2 + \left( \gamma_{\text{PN}} - 1 \right) \Theta(r - \Lambda) \right] \Phi(r), \label{eq:modelsigma}
\end{equation}
where $ r $ and $ \Lambda $ are physical distances, and $ \Theta $ is the Heaviside step function.
In what follows, we mostly make use of the same model as in Ref.~\cite{Jyoti:2019pez}, which is reviewed in Appendix~\ref{app:gravslip}.

The TD between two images of a source at redshift $ z_{S} $ that is gravitationally lensed by a deflector at redshift $ z_{L} $ can be written as
\begin{equation}
\Delta t_{ij} = \frac{1+z_{L}}{c}\frac{D_{L}D_{S}}{D_{LS}}
		\left[ \frac{1}{2} \left( \alpha_{i}^{2} - \alpha_{j}^{2} \right) - \left( \psi_{i} - \psi_{j} \right) \right],
\label{eq:TDij}
\end{equation}
where $ \psi_{i} = \psi(\theta_{i}) $ and $ \alpha_{i} = \alpha(\theta_{i})=\partial_{\theta}\psi|_{\theta_{i}} $ are the \emph{lensing potential} and the \emph{deflection angle} at the angular position (of an image) $ \theta_{i} $, respectively; and $ \{ D_{S},\, D_{L},\, D_{LS} \} $ are the angular diameter distances from us to the source, to the lens and between the lens and source, respectively. Eq.~\eqref{eq:TDij} may be written more compactly in terms of the Fermat potential, $ \phi_{i} =  \alpha_{i}^{2} /2-\psi_{i}$, as
\begin{equation}
\Delta t_{ij} = \frac{D_{\Delta t}}{c}\Delta\phi_{ij},
\label{eq:TDFermat}
\end{equation}
where $ D_{\Delta t} =  (1+z_{L})D_{L}D_{S}/D_{LS} $ is the \emph{time delay distance}.
As the lensing potential depends directly on $ \Sigma $ (see Appendix~\ref{app:gravslip}), applying Eq.~\eqref{eq:modelsigma} results in the decomposition of the lensing potential (and therefore the deflection angle as well) as follows
\begin{equation}
\psi = \psi_{\rm GR} + \left( \gamma_{\text{PN}} - 1 \right)\Delta\psi,
\label{eq:fulllenspotential}
\end{equation}
where $ \psi_{\rm GR} $ is the usual lensing potential in GR, and $ \Delta\psi $ is a lensing potential \emph{slip-term}. 
Assuming a spherical power-law mass density distribution $ \rho_{\rm lens} $, the expressions for the lensing potential components are \cite{Suyu:2012rh,Jyoti:2019pez}
\begin{eqnarray}
\psi_{\rm GR}(\theta) &=& \frac{\theta_{E,\rm GR}^{\gamma'-1}}{3-\gamma'} \theta^{3-\gamma'}, \\
\label{eq:psiGR}
\Delta\psi(\theta)&=&\frac{c\, \theta_{E,\rm GR}^{\gamma'-1}\, }{3-\gamma'} \left(\frac{D_L}{\Lambda}\right)^{\gamma'-3} \\
\label{eq:psislip}
&~&~\times~{}_2F_1\left[\frac{1}{2},\frac{\gamma'-3}{2},\frac{\gamma'-1}{2};\left(\frac{D_L\theta}{\Lambda}\right)^2\right], \nonumber
\end{eqnarray}
where $ \theta_{E,\rm GR} $ is the Einstein radius of the lens, $ \gamma' \equiv - d\log\rho_{\rm lens}/d\log r $ is the radial profile slope, $ {}_2F_1 $ is the hypergeometric function,
\begin{equation}
c = \frac{1 }{2\sqrt{\pi}}~ \frac{\Gamma\left(\frac{\gamma'}{2}-1\right)}{\Gamma\left(\frac{\gamma'-1}{2}\right)},
\end{equation}
and $ \Gamma $ is the Gamma function. The deflection angle components, $ \alpha_{\rm GR} $ and $ \Delta\alpha $, are given by  taking their derivative with respect to $ \theta $.

The gravitational slip correction to the gravitational potential affects the lens parameters, inferred from the lensing observables (observed TD, image positions, etc.). Thus, for an observed lensing event, one should carry out a full Markov-Chain Monte-Carlo (MCMC) analysis for the entire set of parameters (lens parameters plus $ \gamma_{\rm PN} $). 

However, as stongly-lensed FRB systems have not yet been observed, we adopt a more naive approach, suggested in Ref.~\cite{Jyoti:2019pez}, of relating $ \theta_{E,\rm GR} $, the Einstein radius in GR, to $\theta_{E,\rm obs}$, the observed value, which would be inferred differently in the case of screening, via
\begin{equation}
\theta_{E,\rm obs} = \alpha_{\rm GR}(\theta_{E,\rm obs}) +(\gamma_{\rm PN}-1) \Delta\alpha(\theta_{E,\rm obs}),
\label{eq:thetaEcorrection}
\end{equation}
where both $ \alpha_{\rm GR} $ and $  \Delta\alpha $ depend on $ \theta_{E,\rm GR} $ (derived from Eq.~\eqref{eq:psiGR}-\eqref{eq:psislip}). Hence, instead of letting all the lens parameters vary, as one does in a complete MCMC analysis, we account only for the shift in the \emph{observed} Einstein radius.
But again, as there is no $ \theta_{E,\rm obs} $ available for a strongly-lensed FRB system, we must rely on simulation alone. Therefore, in our work we calculate the GR quantities and numerically solve the set of Eqs.~\{\eqref{eq:TDij},\eqref{eq:thetaEcorrection}\} for $ \gamma_{\rm PN} $ and $ \theta_{E,\rm obs} $, where
\begin{equation}
\Delta t_{ij} \rightarrow \Delta t_{\rm obs} + \sigma\left(\Delta t_{\rm obs}\right),
\end{equation}
and
\begin{equation}
\Delta t_{\rm obs} = \frac{D_{\Delta t}}{c}\Delta\phi\left( \theta_{E,\rm obs} \right)
\end{equation}
is the TD one would naively observe in case of  $ \gamma_{\rm PN} \ne 1 $. Note that in case of $  \sigma\left(\Delta t_{\rm obs}\right) = 0 $ the solution yields $ \gamma_{\rm PN} = 1 $ and $ \theta_{E,\rm GR} = \theta_{E,\rm obs} $, thus the deviation from GR is driven by the uncertainties.

\subsection{Uncertainty Contributions}

There are several contributions to the uncertainty in Eq.~\eqref{eq:TDij} \cite{Treu:2016ljm,Treu:2010uj}: (i) the TD measurement error; (ii) the uncertainty in the lens modeling, which results in an uncertainty in the inferred Fermat potential differences; and (iii) the uncertainty in the estimates of the \emph{external convergence}, $ \kappa_{\rm ext} $, which corresponds to the  contribution of the mass distribution along the line of sight (LOS), and results in an overall rescaling of the observed TD.

A typical galaxy-lensing TD is of order $ \sim\!10 $ days. While the relative error in the TD measurement of a lensed quasar is $ \sim 1\% $, the measurement is expected to be extremely accurate in the case of strongly-lensed FRBs due to the short duration of the signal, yielding  relative errors as small as $ \sim\! 10^{-7} \%$, thus completely negligible.

Lens modeling requires high-resolution localization of the lensed FRB images as well as an image of their host galaxy, in order to have better measurements of the angular positions and shear, which in turn result in smaller uncertainty on the Fermat potential differences. Unlike quasars, FRBs have the advantage of not being associated with bright active galactic nuclei, allowing cleaner host images to be obtained (using high-resolution radio telescopes such as VLA or SKA, and VLBI observations). We adopt the relative uncertainty in the Fermat potential differences of $ \delta(\Delta\phi) = 0.8\% $, inferred from a series of simulations of such systems that was recently carried out in Ref.~\cite{Li:2017mek}. We note that the uncertainty in the Fermat potential differences depends, in general, on the system (i.e.\ on the lens and source redshifts, the lens parameters, the sky localization of the event, etc.), and that the uncertainty we use in this work, which we treat as a typical value, was inferred from thousands of simulations of various systems~\cite{Li:2017mek}. In addition, this uncertainty is considered for a single repetition and may be improved by observing multiple repetitions of the same strongly-lensed FRB (by mitigating the lens modeling uncertainty, see Ref.~\cite{Wucknitz:2020spz}). We leave such analysis to future work.

The last and most dominant contribution to the TD uncertainty is due to mass along the LOS. It can be shown that transforming  the lensing potential according to
\begin{equation}
\psi(\theta)\rightarrow\lambda\psi(\theta)+\frac{1-\lambda}{2}\theta^{2},
\end{equation}
together with an isotropic scaling of the \emph{impact parameter} $ \boldsymbol{\beta}\rightarrow\lambda\boldsymbol{\beta} $, results in the same lensing observables (i.e. image positions, magnification ratios, etc.), whereas the TD is shifted by the same factor of $ \lambda $ \cite{1985ApJ...289L...1F,Saha:2000kn,Munoz:2017cll}. This property is known as the \emph{mass-sheet degeneracy} (MSD). Writing the scaling as $ \lambda=1-\kappa_{\rm ext} $, the additional term in the lensing potential can be interpreted as a constant external convergence $ \kappa_{\rm ext} $, which appears due to mass along the LOS. Various techniques, such as using galaxy counts and shear information to obtain the probability density function for $ \kappa_{\rm ext} $, may allow to break this degeneracy and estimate the external convergence and its effect on the TD uncertainty \cite{stx285,Tihhonova:2017mym,Bonvin:2016crt}. As an example, different analyses of HE0435-1223 report values of $ \sigma(\kappa_{\rm ext})=0.013-0.025 $ \cite{Rusu:stx285,Bonvin:2016crt,Tihhonova:2017mym} which correspond to $ \sim1.6-2.5\% $ uncertainty in the TD distance. Below we will follow Ref.~\cite{Li:2017mek}, where the average contribution of the LOS environment to the TD relative uncertainty of lensed FRBs is taken to be $ \delta_{\rm ext}(\Delta t)=0.02 $. We use a single value throughout, although the uncertainty in the external convergence depends on the field of view of the lens and may vary from one system to another.

Therefore, in our analysis we consider the total TD uncertainty to be
\begin{equation}
\sigma(\Delta t) = \Delta t \sqrt{\left[\delta(\Delta\phi)\right]^{2} + \left[\delta_{\rm ext}(\Delta t)\right]^{2}}=\Delta t\,\delta_{\Delta t},
\end{equation}
where $ \delta_{\Delta t}\approx 2.2\% $.

For completeness, it is worth mentioning that there are additional, subdominant, contributions to the uncertainties in the TD measurement, which originate, for example, from the relative motion of the source, lens and the Earth \cite{Dai:2017twh,Zitrin:2018let,Wucknitz:2020spz}, gravitational waves interference \cite{Pearson:2020wxb}, etc. Such effects correspond to $\mathcal{O}(1\,{\rm ms})$ variation in the TD differences and are therefore negligible compared to the contributions we considered above.

\subsection{Simulation}\label{sec:simulation}

In order to simulate strongly-lensed FRB events we must make certain assumptions regarding the distributions of the different parameters of the system. As described in Section~\ref{subsec:model}, in this work we assume spherical symmetry for simplicity, which reduces the number of lens parameters required in more realistic models (e.g. shear, eccentricity, etc.). We follow a three-step procedure to simulate a strongly-lensed FRB system: 1) first we generate the FRB source redshift $ z_{S} $, assuming an FRB density distribution modeled after the newly-released CHIME catalog~\cite{CHIMEFRB:2021srp}, and then determine a corresponding lens redshift $ z_{L} $, 2) next we generate the lens mass using the halo mass function (HMF), thus determining the GR Einstein radius $ \theta_{E,\rm GR} $, 3) finally, we generate the impact parameter, assuming a simple probability distribution, and determine the two images positions $ \theta_{\pm} $.

Throughout our analysis we set the Hubble constant and matter density to $ H_{0} = 67.36 $ km/s/Mpc and $ \Omega_{m} = 0.315 $, respectively, in agreement with best fit values reported by Planck 2018 \cite{Planck:2018vyg}. However, as we discuss below, our results are insensitive to the variation of these parameters within the range suggested by other measurements.
In addition, since observations suggest that early-type lens galaxies have approximately isothermal mass density profiles, i.e.\ $ \gamma' \approx 2 $ \cite{Koopmans:2009av,Ruff:2010rv,Auger:2010va,Barnabe:2011gb,Bolton:2012uh}, we approximate each lens in our simulations as a singular isothermal sphere (SIS). In practice, in order to preserve the angular dependence in Eqs.~\eqref{eq:psiGR}-\eqref{eq:psislip}, we set the radial-profile slope of the lens to $ \gamma' = 1.95 $ (a 2.5\% deviation from the slope of an SIS profile, similar to that of the RXJ1131 lens \cite{Suyu:2012aa}; for more discussion see Appendix~\ref{app:tests}). We note that the SIS approximation is used here as a simplification,  for brevity in the derivations, and yet emphasize that it is valid for the scope of this work, as discussed below.

\subsubsection{FRB and lens redshifts: $ z_{S} $ and $ z_{L} $}
\label{sec:frbdist}

We use the cumulative data, with a total of 669 FRBs, from the CHIME~\cite{CHIMEFRB:2021srp} 
  and FRBcat\footnote{\url{http://www.frbcat.org}}~\cite{Petroff:2016tcr} catalogs, to approximate the redshift distribution of FRB sources (in previous works the distribution was assumed to follow the star-formation history or to have a constant comoving density \cite{Munoz:2016tmg, Li:2017mek}, but these no longer fit the data  well). 
  
In order to infer the redshift of each FRB from the observed DM, we follow the formalism in Ref.~\cite{James:2021jbo}, and model the DM as
\begin{equation}
{\rm DM} = {\rm DM}_{\rm ISM} + {\rm DM}_{\rm halo} + {\rm DM}_{\rm cosmic} + {\rm DM}_{\rm host},
\label{eq:dmmodel}
\end{equation}
where the first two terms are the contributions from the inter-stellar medium (ISM) and the dark-matter halo of the Milky Way, and the last two terms are the \emph{extra-galactic} contribution, composed of the DM due the source and its host galaxy, and the \emph{cosmic} DM, which accounts for the contribution from the \emph{intergalactic medium} (IGM), given by
\begin{eqnarray}
{\rm DM}_{\rm cosmic}(z) &=& \int_{0}^{z}\frac{c \bar{n}_{e}(z')}{1+z}\left| \frac{dt}{dz'}\right| dz',\label{dmcos}\\
&=& \int_{0}^{z}\frac{c \bar{n}_{e}(z')dz'}{H_{0}(1+z)^{2}\sqrt{\Omega_{m}(1+z)^{3}+\Omega_{\Lambda}}},\nonumber
\end{eqnarray}
\begin{figure}[htbp!]
\includegraphics[width=0.49\textwidth]{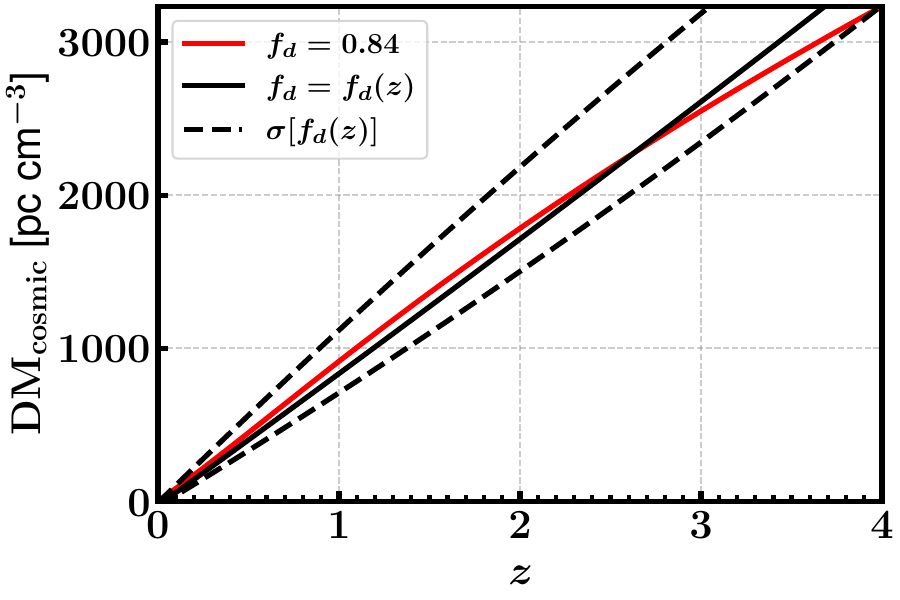}
\caption{The redshift dependence of $ {\rm DM}_{\rm cosmic} $. In solid red we show our results, evaluated by assuming a constant $ f_{d}=0.84 $ in Eq.~\eqref{dmcos}; in solid black we show the results from Ref.~\cite{James:2021jbo}, where $ f_{d} $ is modeled to be redshift dependent, and its uncertainty in dashed black. We find that the results are well correlated in the range  $ 0<z\lesssim4 $.}
\label{fig:dmzrelation}
\end{figure}
with $ \bar{n}_{e}(z) = f_{d}\rho_{b}(z)m_{p}^{-1}(1-Y_{\rm He}/2) $, where $ Y_{\rm He} = 0.25 $ is the Helium mass fraction; $ m_{p} $ is the proton mass; $ \rho_{b} = \Omega_{b}\rho_{c,0}(1+z)^{3} $ is the baryon density in units of $ \rho_{c,0}=3H_{0}^{2}/8\pi G $, for which we assume $ \Omega_{b}h^{2}=0.0224 $ as reported by Planck \cite{Planck:2018vyg}; and $ f_{d} $ is the fraction of baryons in the IGM. The baryon fraction $ f_{d} $ is, in general, an evolving quantity, which can be modeled as done in Ref.~\cite{Li:2019klc,Macquart:2020lln,James:2021jbo}. However, for the purpose of this work we follow Ref.~\cite{Zhang:2020ass} and use a constant value of $ f_{d}=0.84 $, for which the resulting $ {\rm DM}_{\rm cosmic}(z) $ is consistent with the one in Ref.~\cite{James:2021jbo}, for $ z\lesssim4 $, as we show in Fig.~\ref{fig:dmzrelation}.
In our analysis we make use of the excess DM given in the catalogs, where the ISM contribution, estimated via NE2001 model \cite{Cordes:2002wz}, had been already subtracted from the observed DM. The DM contribution of the Milky Way's halo and the host are uncertain, we thus follow the assumptions in Ref.~\cite{James:2021jbo} and use the mid-range value of ${\rm DM}_{\rm halo}=50\text{ pc cm}^{-3} $, and the best-fit value from the analysis in Ref.~\cite{James:2021jbo}, $ {\rm DM}_{\rm host} = 145 \text{ pc cm}^{-3}/(1+z) $, which is weighted by the redshift of the source.
\begin{figure}[htbp!]
\includegraphics[width=0.49\textwidth]{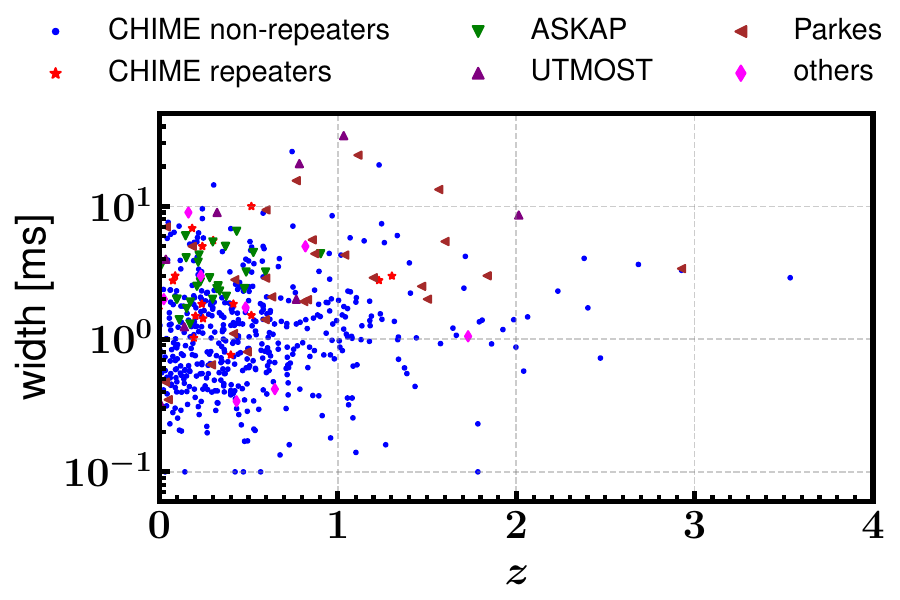}
\caption{Redshift distribution of 669 FRBs from the CHIME catalog~\cite{CHIMEFRB:2021srp} and FRBcat~\cite{Petroff:2016tcr}, where \emph{others} denotes FRBs detected by GBT, VLA, Apertif, DSA-10, FAST and Arecibo.}
\label{fig:frbscatter}
\end{figure}
We estimate the redshifts of the detected FRBs by solving Eq.~\eqref{eq:dmmodel} for $ z $, and plot their distribution in Fig.~\ref{fig:frbscatter}. We note that in our analysis we neglect the uncertainties in the formalism above, which should be accounted for when running a full MCMC analysis, as done in Ref.~\cite{James:2021jbo}.
We find that the FRB redshift distribution can be described to a good approximation by the simple distribution
\begin{equation}
N_{\rm FRB} (z) = \frac{z}{z_{c}^{2}}e^{-z/z_{c}},
\label{eq:frbdistfunction}
\end{equation}
where the parameter $ z_{c} =\langle z \rangle/2 = 0.255 $ sets the expected value of $ z $. In Fig.~\ref{fig:frbdist} we show the histogram of the combined catalogs, normalized to unity, along with the best-fit distribution function. As using data from different experiments may introduce a selection bias into fit,  we repeated the fit using the CHIME catalog alone, which yields  $ z_{c} =\langle z \rangle/2 = 0.256 $, corresponding to a sub-precent deviation in the constraining power on $ \gamma_{\rm PN}$.
\begin{figure}[htbp!]
\includegraphics[width=0.49\textwidth]{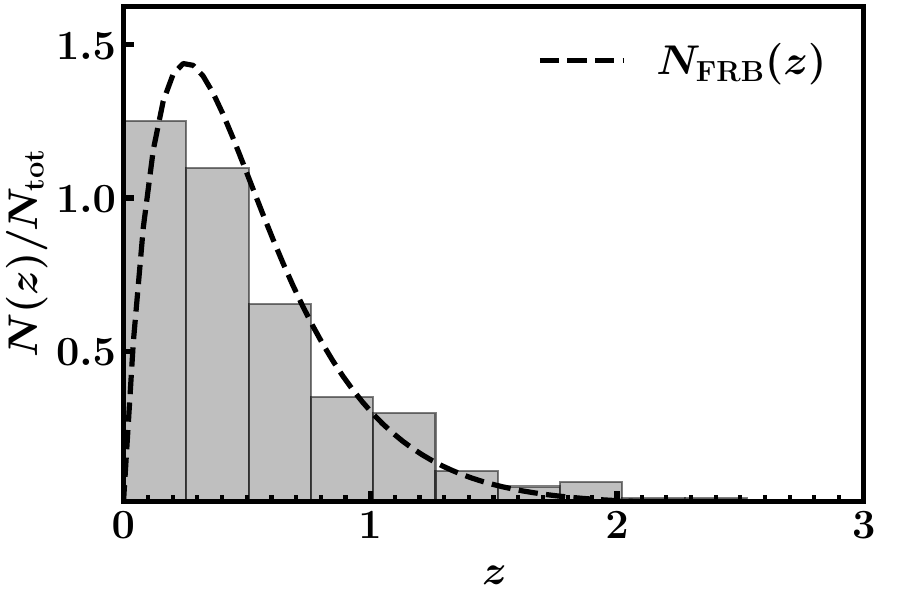}
\caption{A histogram of  669 FRBs from the CHIME catalog~\cite{CHIMEFRB:2021srp} and FRBcat~\cite{Petroff:2016tcr}, normalized to unity. The assumed FRB redshift distribution in Eq.~\eqref{eq:frbdistfunction} is plotted in a dashed line. The distribution fits the data well, however it tends to slightly overestimate (underestimate) lower (higher) redshifts.}
\label{fig:frbdist}
\end{figure}

In our analysis we generate a redshift for each FRB independently, according to Eq.~\eqref{eq:frbdistfunction}.
We then repeat the choice of Ref.~\cite{Li:2017mek} and use the source redshift to determine the corresponding lens redshift $ z_{L} $, by setting it to be the value which maximizes the probability for a distant point source at redshift $ z_{S} $ to be lensed by an intervening DM halo at $ z_{L}$, as presented in Fig.~2 of Ref.~\cite{Li:2014dea}.

\subsubsection{Einstein radius $ \theta_{E,\rm GR} $}

The Einstein radius $ \theta_{E,\rm GR} $ is determined by the lens parameters. In particular, for a SIS, it is given by \cite{2017grle.book.....D}
\begin{equation}
\theta_{E,\rm GR} = \sqrt{\frac{4GM(\theta_{E,\rm GR})}{c^{2}}\frac{D_{LS}}{D_{L}D_{S}}},
\label{eq:thetaEGR}
\end{equation}
where $ G $ is  Newton's constant and $ M(\theta_{E,\rm GR}) $ is the mass enclosed within a radius of $ \theta_{E,\rm GR} $. Therefore, in order to set the Einstein radius of a lens at redshift $ z_{L} $ we must specify its mass, which we assume to follow the HMF within the  range  $ 10^{10}h^{-1}M_{\odot} \!<\! M \!< \!2\times 10^{13}h^{-1}M_{\odot} $ ($ H_{0}=h\times100 $ km/s/Mpc), which roughly corresponds to dark-matter halos with this profile that reside in galaxies and make the dominant contribution to the lensing probability \cite{Navarro:1996gj,Li:2000dw,Li:2002ec,Li:2014dea}.
We consider the probability density for lensing by a dark-matter halo with mass $ M $, at redshift $ z_{L} $, to be proportional to $ n(M,z)\sigma_{\rm SIS}(M) $, where $ n(M,z) $ is the number density of halos (determined by the HMF) and $ \sigma(M)=\pi\theta_{E}^{2} $ is the SIS lensing cross-section, taken to be the cross-section of its Einstein sphere.
We make use of the publicly available HMF code\footnote{\url{https://github.com/halomod/hmf}} \cite{2013A&C.....3...23M}, assuming the Press-Schechter formalism \cite{Press:1973iz}, to generate the HMF at a given lens redshift $ z_{L} $, within the aforementioned mass range. We then use the HMF as the distribution for generating the lens mass, which determines $ \theta_{E,\rm GR} $.

\subsubsection{Impact parameter $ \beta $}

We assume that the probability for a source at angular position $ \beta $ (the impact parameter), in the projected source plane, to be lensed by an intervening halo centered at the origin (at the LOS), is proportional to the circumference of a ring with radius $ \beta $. Thus, we randomly choose the impact parameter from a distribution that is linear with $ \beta $ and has a cutoff at $ 2\theta_{E,\rm GR}/3 $ (set by a limit on the flux ratio of the image, as described in Appendix~\ref{app:simmodel}). We also exclude values smaller than a critical value set by the limit on FRB TD detection, $ \beta\gtrsim c\Delta t_{\rm min}/(2D_{\Delta t}\theta_{E}) $, which is of order $ \sim10^{-9}\,{\rm arcsec} $ and is therefore practically negligible (the probability for such value vanishes).
Given the Einstein radius and impact parameter we then determine the angular position of the images according to the SIS lens model, $ \left|\theta_{\pm}\right| = \theta_{E,\rm GR} \pm \beta $.
A more detailed description can be found in Appendix~\ref{app:simmodel}.

\section{Results and Discussion}\label{sec:results}

\begin{figure}[htbp!]
\includegraphics[width=0.49\textwidth]{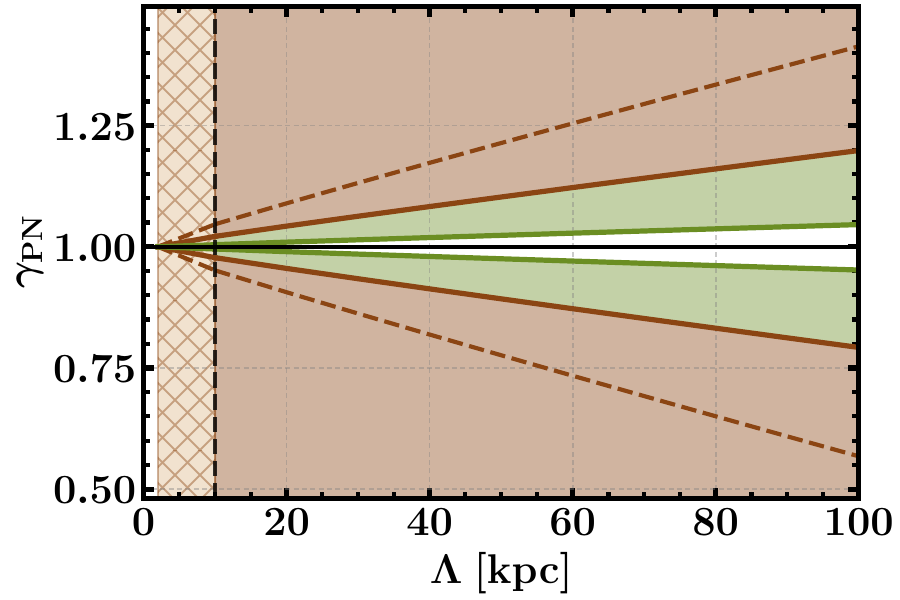}
\caption{The constraints on $ \gamma_{\rm PN} $ from 10 simulated event (solid green), single average event (solid brown) and its $ 1\sigma $ uncertainty (dashed brown). Shaded areas are excluded by the constraints of the corresponding color. A black vertical dashed line signifies the $10\,{\rm kpc}$ threshold.}
\label{fig:mainresult}
\end{figure}

We use the procedure described in Section~\ref{sec:simulation} to simulate each strongly-lensed FRB, then solve Eqs.~\{\eqref{eq:TDij},\eqref{eq:thetaEcorrection}\} to find the constraints  it can place on $ \gamma_{\rm PN} $ for each screening radius $ \Lambda $. We simulate a modest number of $ N\!=\!10 $ events, which we assume can be expected to be observed in the next few years, averaged over 100 simulations, to eliminate statistical variance. In Fig.~\ref{fig:mainresult} we show the upper and lower constraints on $ \gamma_{\rm PN} $ at $1\sigma$ C.L. from an average single event; and from the combination of 10 events. The range of screening radii displayed in Fig.~\ref{fig:mainresult} was chosen to be the same as in Ref.~\cite{Jyoti:2019pez}, setting the minimal screening radius at a typical Einstein radius value, $ \Lambda \gtrsim \theta_{E}D_{L}\approx 10\rm kpc $.
However, as the assumption of an abrupt transition at radius $ \Lambda $ is an approximated simplification to the real screening predicted by models, our results are expected to be less accurate in the proximity of the lower limit.

\begin{figure}[b!]
\includegraphics[width=0.46\textwidth]{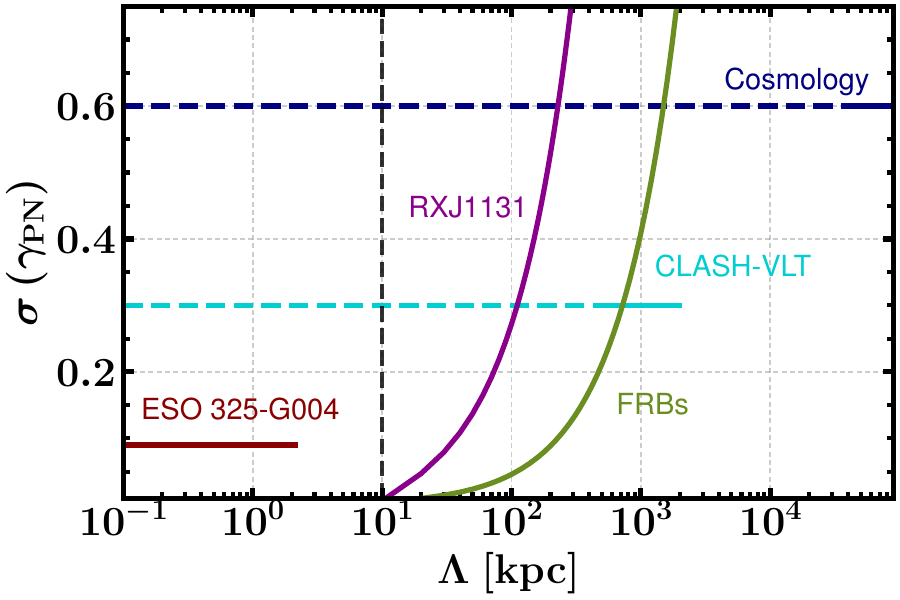}
\caption{1-$ \sigma $ uncertainty on $ \gamma_{\rm PN} $ as a function of the screening radius $ \Lambda $. We show our prediction for the constraining power of TDs from 10 strongly-lensed FRBs (solid green) in comparison to constraints obtained from:
a nearby strong gravitational lens ESO 325-G004 \cite{Collett:2018gpf} (solid red), where the cutoff corresponds to the Einstein radius of the lens $ \Lambda<\theta_{E}D_{L} $;
clusters observations by CLASH and VLT \cite{Pizzuti:2016ouw} (solid cyan);
the cosmological observational data from Planck and large scale structure \cite{Planck:2015bue} (solid blue);
and the constraints from TD of a strongly-lensed quasar by RXJ1131 \cite{Jyoti:2019pez} (solid purple).
Since all of these measurements place constant constraints on $ \gamma_{\rm PN} $ at a certain distance scale, we also plot dashed lines to represent the upper limit for the constraints on smaller scales.}
\label{fig:comparison}
\end{figure}

The constraining power of all 10 events is evaluated according to the combined variance around the fiducial value $ \gamma_{PN} = 1 $
\begin{equation}
\left(\gamma_{PN}^{\pm}\right)_{\rm total} = 1 \pm \left[\sum_{i=1}^{n=10}\frac{1}{\sigma_{\pm,i}^{2}}\right]^{-1/2},
\end{equation}
where $ \sigma_{\pm,i}^{2} = \left| (\gamma_{PN}^{\pm})_{i} - 1 \right|^{2} $ is the variance from the $ i $-th event (the $ \pm $ corresponds to the maximum and minimum allowed values for $ \gamma_{\rm PN} $). We find that the combined constraint on $ \gamma_{\rm PN} $ from $ N\!=\!10 $ events scales as $ \left| \gamma_{\rm PN}-1\right| \lesssim 0.04\times(\Lambda/100\rm kpc)\times[N/10]^{-1/2} $ (for $ \Lambda\lesssim100 $ kpc), which is roughly 3 times stronger than the corresponding constraint from 10 strongly-lensed quasars, estimated from Ref.~\cite{Jyoti:2019pez}. In previous work~\cite{Bolton:2006yz,Schwab:2009nz,Smith:2009fn,Pizzuti:2016ouw,Collett:2018gpf}, that searched for MG by studying strongly-lensed systems, the constraints were inferred from analyzing the distance-dependent deviation from GR, hence they are sensitive to  scales smaller than the Einstein radius, $ \Lambda\lesssim\theta_{E} D_{L} $ (e.g.\ $ \sim\!{\rm kpc} $ for galaxies and $ \sim\!{\rm Mpc} $ for clusters).
A comparison between the constraining power we predict from near future detections of strongly-lensed FRBs to those from previous work is shown in Fig.~\ref{fig:comparison}. We find that observations of FRB TDs will allow a significant increase of the constraining power on the Post-Newtonian parameter $ \gamma_{\rm PN} $ in the regime of screening radii $ \Lambda = 10-800\rm kpc $.

We have tested the robustness of our results against variation in the cosmological parameters, $ H_{0} $ and $ \Omega_{m} $, within the  ranges suggested by recent experimental constraints such as Planck \cite{Planck:2018vyg} and local measurements of $ H_{0} $ \cite{Riess:2019cxk}, and found a negligible impact of $ \lesssim0.1\% $ on the constraining power on $ \gamma_{PN} $, for the set of parameters specified in Appendix~\ref{app:tests}, where we demonstrate the responsiveness of our constraints to the different lens parameters. We have also tested the impact of the relative uncertainty and found that the constraints on $ \gamma_{\rm PN} $ vary linearly with $ \delta_{\Delta t} $ for a given screening radius, and scale as $ \left| \gamma_{\rm PN}-1\right| \lesssim 0.12\times(\Lambda/100\rm kpc)\times[N/10]^{-1/2} $ for $ N\!=\!10 $ events, if we increase the relative uncertainty to $ \delta_{\Delta t} \!=\! 6\% $. 

In addition, we made a consistency check of our SIS profile ansatz, repeating the analysis above when simulating the systems using the exact lens equations for the power-law mass profile \cite{Suyu:2012rh}, reviewed in Appendix~\ref{app:powerlawmodel}, and found that our SIS approximation holds up to $ \sim0.1\!-\!0.6\% $ precision in the constraints for $ \Lambda\!=\!100-800 $ kpc, and $ \lesssim 0.1\% $ for $ \Lambda<100 $ kpc, justifying our approximations.

\section{Conclusions}\label{sec:conclusions}

In this work we have studied the application of strongly-lensed (repeating) FRBs to set constraints on a wide range of MG models, which exhibit a gravitational slip, parametrized by $ \gamma_{\rm PN} $, beyond a given screening radius $ \Lambda $. 
We used the results of recent work concerning FRBs to simulate a set of strongly-lensed FRB systems.
We then used the phenomenological model from Ref.~\cite{Jyoti:2019pez}, along with recent estimations of the relative time-delay uncertainties for such systems, to simulate a set of observed events and estimate the maximal and minimal values of $\gamma_{\rm PN}$ that are consistent with them.

While current surveys have already catalogued a large number of FRBs~\cite{CHIMEFRB:2021srp}, upcoming surveys will vastly expand this catalogue~\cite{5136190,Newburgh:2016mwi,PUMA:2019jwd}. Once the observed number approaches $ \sim 10^{4} $ it is likely that a few will prove to be strongly-lensed repeaters, for which a full MCMC analysis could be used in order to constrain all the lens parameters as well as $ \gamma_{\rm PN} $. We have shown that this can improve the constraining power on MG, in terms of both precision and  range of screening radii, as 10 events alone could place constraints at a level of 10\% in the range of $ \Lambda = 10 - 300 $ kpc and to improve the precision of current constraints up to distances of $ \sim800 $ kpc.
Furthermore, due to their high-precision TD measurement, strongly-lensed repeating FRBs may offer the opportunity for observing ``real time'' evolution of the lens by studying the change in the TD over time. Such observable can serve as a new probe of cosmic expansion \cite{Zitrin:2018let} as well as for mitigating uncertainties from the mass distribution of the lens, as proposed and studied in Ref.~\cite{Wucknitz:2020spz}. This can improve the constraints from TDs from strongly-lensed FRB on screened MG theories with  gravitational slip.

\acknowledgements

We thank Julian B. Munoz for useful discussions and helpful comments on an earlier version of this manuscript.
We also thank the anonymous referee for comments which helped improve the clarity of this manuscript.
EDK is supported by a faculty fellowship from the Azrieli Foundation. TA is supported by a Negev PhD fellowship.


\bibliography{refs}

\appendix

\section{Lensing Potential}\label{app:gravslip}

The lensing potential is defined as \cite{Narayan:1996ba}
\begin{equation}
\psi(\theta) = \frac{1}{c^{2}} \frac{D_{LS}}{D_{L}D_{S}} \int \Sigma(D_{L} \theta, z) dz,
\end{equation}
where $ z $ denotes the distance along the line of sight (LOS) and $ \Sigma $ is the sum of the potentials defined in Eq.~\eqref{eq:modelsigma}, which reduces to $ 2\Phi $ in GR. Thus, the two contributions to the lensing potential, noted in Eq.~\eqref{eq:fulllenspotential} are
\begin{eqnarray}
\psi_{\rm GR}(\theta) &=& \frac{2}{c^{2}} \frac{D_{LS}}{D_{L}D_{S}} \int \Phi(D_{L} \theta, z) dz,\\
\Delta\psi(\theta) &=& \frac{1}{c^{2}} \frac{D_{LS}}{D_{L}D_{S}} \int \Theta(r-\Lambda)\Phi(D_{L} \theta, z) dz.
\end{eqnarray}
Assuming a power-law mass profile of the form
\begin{equation}
\rho(r) = \rho_{0} \left(\frac{r_{0}}{r}\right)^{\gamma'},
\label{eq:massprofile}
\end{equation}
where $ \rho_{0} $ and $ r_{0} $ are constant parameters which set the mass of the lens. Employing the Poisson equation, $ \nabla^{2}\Phi = 4\pi G a^{2} \rho $, yields the lens Newtonian potential 
\begin{equation}
\Phi(r) = \frac{4\pi G \rho_{0} r_{0}^{\gamma'}}{(3-\gamma')(2-\gamma')}r^{2-\gamma'},
\end{equation}
which leads to Eqs.~\eqref{eq:psiGR}-\eqref{eq:psislip}, where $ \theta_{E,\rm GR} $ encompasses the coefficients of the mass profile.
The deflection angle can be derived from the lensing potential via $ \bs{\alpha}(\bs{\theta}) = \nabla_{\bs{\theta}}\psi(\bs{\theta}) $, leading to the two contributions
\begin{eqnarray}
\alpha_{\rm GR}(\theta) &=& \partial_{\theta}\psi_{\rm GR}(\theta)\\
\Delta\alpha(\theta) &=& \partial_{\theta}\Delta\psi(\theta).
\end{eqnarray}

\section{Simulated Lens Modeling}\label{app:simmodel}

The dimensionless lens equation that corresponds to the power-law mass profile \eqref{eq:massprofile} is \eqref{eq:lenseq}
\begin{equation}
\bs{y} = \bs{x} - |x|^{2-\gamma'}\frac{\bs{x}	}{|x|}.
\end{equation}
But, as mentioned in Section~\ref{sec:simulation}, the type of lenses we consider  can be approximated to have a SIS profile, i.e.\ $ \gamma'\approx2 $, reducing the lens equation to the traditional form
\begin{equation}
y = x - \frac{x}{|x|},
\end{equation}
where we dropped the vector notation due to the spherical symmetry and the Einstein radius is defined by Eq.~\eqref{eq:thetaEGR}. It is easy to show that this equation has two solutions (images), $ \theta_{\pm}=\beta\pm\theta_{E} $, only if $ y<1 $, which places an upper boundary on the possible value of the impact parameter ($ \beta<\theta_{E} $). However, a more restrictive boundary can be placed by considering the detection limitations of both events. Following Ref.~\cite{Munoz:2016tmg}, the image flux ratio  is defined as the ratio of the image magnifications $ \mu_{\pm} $,
\begin{equation}
R_{f} = \left|\frac{\mu_{+}}{\mu_{-}} \right|= \frac{1/y+1}{1/y-1} = \frac{1+y}{1-y}>1.
\end{equation}
In order to ensure both events are detected (assuming that the brighter first image is detectable) we set a minimum threshold value $ \bar{R}_{f} $, which then yields an upper limit $y < (\bar{R}_{f} - 1)/(\bar{R}_{f} + 1)$.
Requiring a conservative redshift-independent threshold of $ \bar{R}_{f} = 5$ (see Ref.~\cite{Munoz:2016tmg}) yields a more restrictive boundary of $ \beta<2\theta_{E}/3 $.

The lower limit for the impact parameter is determined by the threshold for detecting the TD between the two images, which must be larger than the duration of the FRB signal. However, since a SIS lens TD is independent of the impact parameter (it is easy to show that it cancels out), we consider the non-approximated expression for the TD of the power-law profile in Eq.~\eqref{eq:massprofile} \cite{Suyu:2012rh}
\begin{equation}
\Delta t=\frac{D_{\Delta t}}{c}\frac{\gamma'-1}{2(3-\gamma')}\left[ \theta_{+}^{2} - \theta_{-}^{2} + \frac{4(2-\gamma')}{\gamma'-1} \beta (\theta_{+} + \theta_{-}) \right].
\label{eq:TDpowerlaw}
\end{equation}
Now, taking the limit of $ \gamma\rightarrow 2 $ and the solutions for the SIS lens equation, Eq.~\eqref{eq:TDpowerlaw} can be approximated as
\begin{equation}
\Delta t \approx 2\frac{D_{\Delta t}\theta_{E}}{c}\beta,
\end{equation}
setting the lower threshold for the impact parameter as
\begin{equation}
\beta\gtrsim 10^{-7}\Delta t_{\rm min}\left[ \frac{\rm Gpc}{D_{\Delta t}}\right]\left[\frac{\rm arcsec}{\theta_{E}} \right]\;[\rm arcsec].
\end{equation}
However, since FRBs have durations of milliseconds, this value is extremely small ($ \sim10^{-9} $arcsec), compared to traditional quasar lensing ($ \Delta t_{\rm min}\sim \rm days $), for which it is $ \sim0.1\,{\rm arcsec}$.

\section{Responsiveness to Lens Parametes}\label{app:tests}

\begin{figure*}[htbp!]
\includegraphics[width=0.98\textwidth]{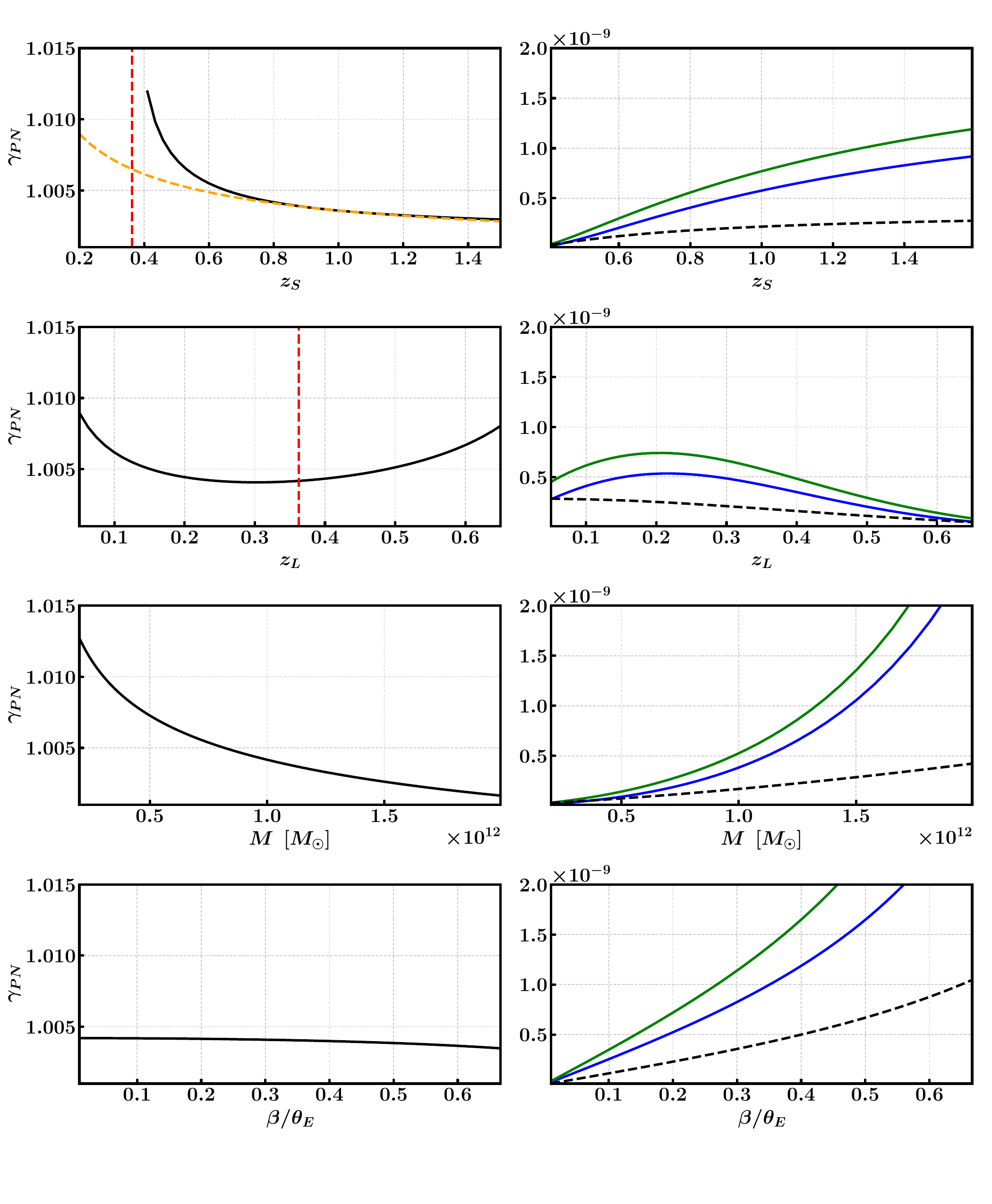}
\caption{The responsiveness in the constraining power on $ \gamma_{\rm PN} $ to variation of the different parameters of the strongly-lensed FRB system (on the left side), and the corresponding responsiveness of the Fermat potential term in Eq.~\eqref{eq:reducedeqn} (on the right side): $ \Delta\times10^{3} $ (dashed black), $ \Delta\phi_{\rm slip}^{(1)}\times10 $ (solid blue) and $ \Delta\phi_{\rm slip}^{(2)} $ (solid green). The dashed red vertical lines represent $ z_{L}=0.36 $, which is associated with $ z_{S}=0.8 $ according to the maximal lensing probability relation. In order to disentangle the redshift responsiveness due to the relative distance between the lens and source, we also plot the responsiveness to the $ z_{S} $ with a dynamical lens redshift (dashed orange), where $ z_{L} $ is determined by the same relation. As implied from Eq.~\eqref{eq:reducedeqn}, the constraining power on $ \gamma_{\rm PN} $ is determined by the ratio between the slip terms, $ \Delta\phi_{\rm slip} $, and $ \Delta $.} 
\label{fig:varyingparams}
\end{figure*}

\begin{figure}[htbp!]
\vspace{3mm}
\includegraphics[width=0.49\textwidth]{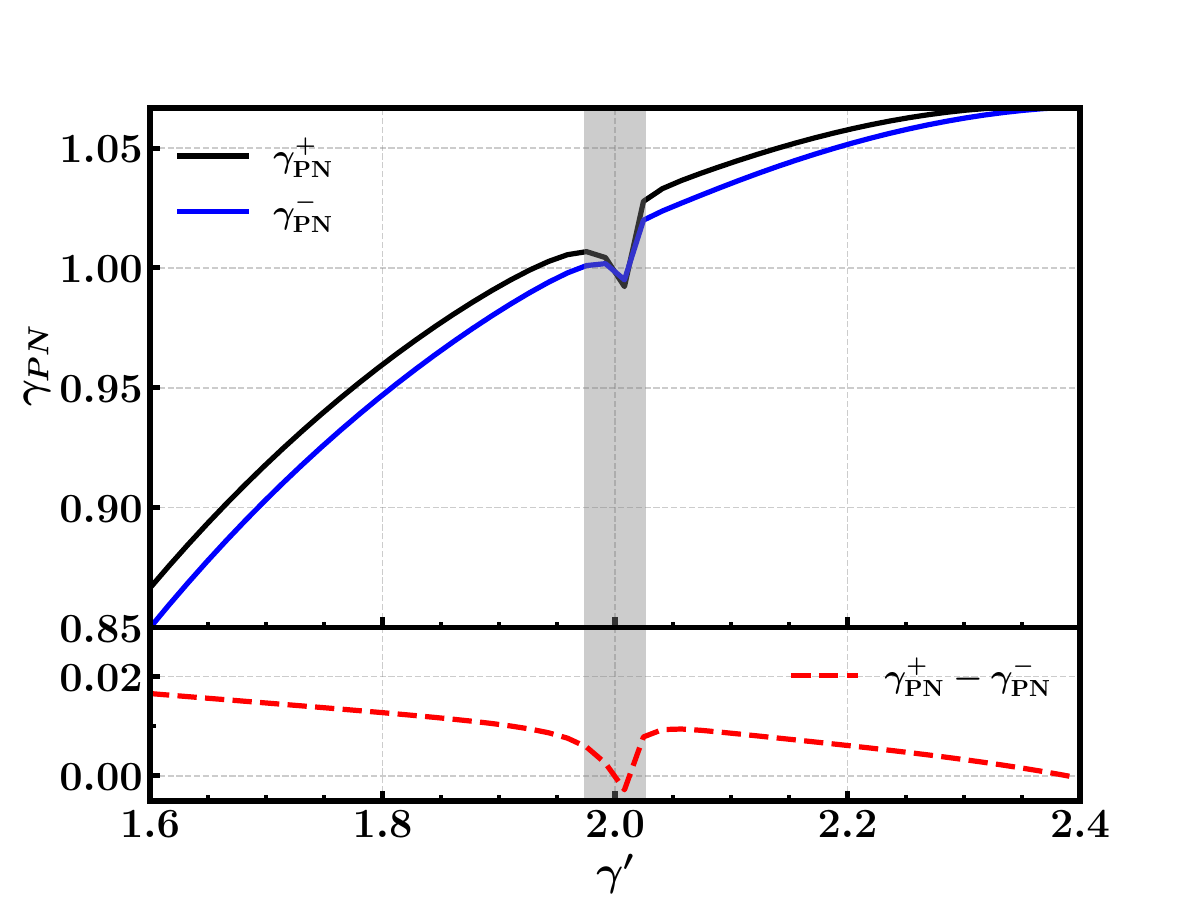}
\caption{The responsiveness of the constraining power on $ \gamma_{\rm PN} $ to a variation of the mass profile slope $ \gamma' $. The gray band around $ \gamma'=2 $ corresponds to the anomalous region where one of the slip terms diverges, which should be excluded to avoid artificially tight constraints. It is clear that there is a degeneracy between the upper and lower constraints on $ \gamma_{\rm {PN}} $ and the slope, resulting in a tilt around the value of $ \gamma_{\rm PN}=1 $. It is also notable that the constraining power, evaluated by the difference $ \gamma_{\rm PN}^{+}- \gamma_{\rm PN}^{-} $, is almost unaffected by the variation of $ \gamma' $, a change of $ \lesssim 1\% $ between $ 1.8 \ge \gamma' \le2.2 $. }
\label{fig:varyingslope}
\end{figure}

In order to make sure our results are robust and consistent, we test the responsiveness of the constraining power to the slope of the lens mass profile $ \gamma' $ in Fig.~\ref{fig:varyingslope}. We also provide in Fig.~\ref{fig:varyingparams} the responsiveness of $ \gamma_{\rm PN} $ to the different parameters we simulated, $ \left\{z_{S},\, z_{L},\, M,\, \beta\right\} $. In these tests we set all parameters to a typical fixed set of values, $ \{ z_{S}=0.8,\, z_{L}=0.36,\, M=10^{12}M_{\odot},\,\beta=\theta_{E}/3,\,\gamma_{\rm PN}=1.95 \} $, while varying only one of them at a time. Since, in different MG theories, the screening radii are usually determined by the geometric mean of the Compton wavelength of the graviton and the Schwarzschild radius of the massive object \cite{Babichev:2013usa}, we choose to perform the tests at  screening radii of $ \Lambda=20 $ kpc, which corresponds to a graviton with Hubble-radius wavelength and a typical galaxy of mass $ \sim10^{12}M_{\odot} $.

The equation for the TD that we solve, Eq.~\eqref{eq:TDij}, can be simplified in our analysis as  both sides are proportional to $ D_{\Delta t}/c $, so that it can be written as
\begin{eqnarray}
\Delta\phi\left(\theta_{E,\rm obs}\right)[1\pm\delta_{\Delta t}] &=& \Delta\phi\left(\theta_{E,\rm GR}\right) +  \left(\gamma_{\rm PN}-1\right)\Delta\phi_{\rm slip}^{(1)} \nonumber \\
&~&~ + \left(\gamma_{\rm PN}-1\right)^{2}\Delta\phi_{\rm slip}^{(2)} ,
\label{eq:reducedeqn}
\end{eqnarray}
where we defined
\begin{eqnarray}
\Delta\phi_{\rm slip}^{(1)} &=& \alpha_{i}\Delta\alpha_{j} - \Delta\psi_{i}+\Delta\psi_{j}, \\
\Delta\phi_{\rm slip}^{(2)} &=& \frac{1}{2}\left(\Delta\alpha_{i}^{2} - \Delta\alpha_{j}^{2} \right),
\end{eqnarray}
which we call the Fermat potential \emph{slip} terms. In what follows we will shift the first term Eq.~\eqref{eq:reducedeqn} to the LHS of the equation, denoting the sum of all terms on that side by $ \Delta $, so that we may refer to three quantities: $ \{\Delta,\, \Delta\phi_{\rm slip}^{(1)},\, \Delta\phi_{\rm slip}^{(2)} \}$. We plot the impact of each parameter in Figs.~\ref{fig:varyingparams} and~\ref{fig:varyingslope}.

It is clear from the bottom plot on the left in Fig.~\ref{fig:varyingparams} that the influence of the impact parameter is negligible since all the Fermat potential terms (the plots to right of Fig.~\ref{fig:varyingparams}) increase linearly with $ \beta $ which prevents $ \gamma_{\rm PN} $ from varying, therefore, the impact of varying other parameters is not influenced by the fact that we take $ \beta=\theta_{E}/3 $ (which naturally varies with $ \theta_{E} $).
We find that as the source is closer to the lens (i.e. smaller $ D_{LS} $), the constraining power on $ \gamma_{\rm PN} $ drops, whereas increasing $ z_{S} $ beyond the value which corresponds to the maximal lensing probability relation does not lead to tighter constraints, as can be seen by looking at the difference between the curves of constant and dynamic $ z_{L} $, at the top left plot in Figure~\ref{fig:varyingparams}.
Considering the low likelihood of detecting events with small $ D_{LS} $, we deduce that the responsiveness to the relative distance between the lens and the source is small. However, when we let $ z_{L} $ vary with $ z_{S} $, according to the maximal lensing probability relation, we find that the the constraining power on $ \gamma_{\rm PN} $ increases with the source redshift.
Nonetheless, higher redshifts yield smaller lens mass, for which the constraining power is smaller.
We also note that the responsiveness of $ \gamma_{\rm PN} $ to the lens redshift (at a given source redshift) features a minimum near the value of maximum lensing probability (vertical dashed red line), due to the dependence of the slip terms on the lens distance, $ D_{L} $, which confirms our assumption of considering the one-to-one relation between $ z_{S} $ and $ z_{L} $ in Sec.~\ref{sec:frbdist}, as the variation of $ z_{L} $ would yield a small impact on the constraints on $ \gamma_{\rm PN} $.

Finally we address the variation of the slope of the power-law mass profile, $ \gamma' $, presented in Fig.~\ref{fig:varyingslope}. The variation of the slope poses a difficulty in our analysis, as the most probable value of $ \gamma'=2 $ results in the diversion of the slip terms - yielding an artificially tight constraint on $ \gamma_{\rm PN} $, whereas values away from 2 are in a mismatch with the simulated SIS values we approximated - mostly the position of the images. We also note the degeneracy between $ \gamma_{\rm PN} $ and the slope, which originates from the degeneracy between the Fermat potentials and the slope. However, as shown in Fig.~\ref{fig:varyingslope}, the tilt has a small impact on the magnitude of the constraints ($ \sim 1\% $ around $ \gamma'=2 $).
Therefore we find that as long as $ \gamma' $ is not too close to the critical value of 2 (and we are outside the grayed region in Fig.~\ref{fig:varyingslope}) or too far---so that the SIS approximation still holds (15\% deviation from $ \gamma' =2$ corresponds roughly to a maximum 10\% deviation in the position of the images)---the inferred constraining power remains valid. In our analysis we used a value of $ \gamma' = 1.95 $---similar to the value of the RXJ1131 lens~\cite{Suyu:2012aa}---for which our SIS approximation holds, the results are almost perfectly symmetric around 1, and we are outside the anomalous region around $ \gamma'=2 $.

\section{Power-Law Mass Profile Lens Model}\label{app:powerlawmodel}

The \emph{surface mass density}, $ \Sigma(R) $ (not to be confused with the sum of potentials $ \Sigma=\Phi+\Psi $), of a spherical power-law mass profile \eqref{eq:massprofile} is given by integrating the mass density along the LOS,
\begin{eqnarray}
\Sigma(R) &=& \int_{-\infty}^{\infty}\rho\left(\sqrt{R^{2}+z^{2}}\right)dz\\
&=&\rho_{0}r_{0}\sqrt{\pi}\left(\frac{r_{0}}{R}\right)^{\gamma-1}\frac{\Gamma\left[\frac{\gamma'-1}{2}\right]}{\Gamma[\gamma'/2]},\nonumber
\end{eqnarray}
where $ R = \theta D_{L} $ is the projected radius on the lens plane, $ dz $ is a distance segment along the LOS (not to be confused with redshift), and we assume $ \gamma>1 $. Using the usual definition of the critical surface density, $ \Sigma_{cr} = c^{2}D_{S}/4\pi G D_{L} D_{LS}$, the dimensionless surface mass density (also known as the \emph{convergence}) is
\begin{equation}
\kappa(\theta) = \Sigma(\theta D_{L})/\Sigma_{cr},
\end{equation}
which can be written more conveniently in terms of the Einstein radius $ \theta_{E} $ as
\begin{equation}
\kappa(\theta) = \frac{3-\gamma'}{2}\left(\frac{\theta_{E}}{\theta}\right)^{\gamma'-1}.
\end{equation}
The relation between the convergence and lensing potential \cite{2006glsw.conf.....M}, $ \nabla^{2}\psi(\theta) = 2\kappa(\theta)$, yields
\begin{equation}
\psi(\theta) = \frac{\theta_{E}^{2}}{3-\gamma'}\left(\frac{\theta_{E}}{\theta}\right)^{\gamma'-3},
\end{equation}
which leads to the deflection angle
\begin{equation}
\bs{\alpha} (\bs{\theta})=\bs{\nabla_{\theta}}\psi(\bs{\theta}) = \theta_{E}^{\gamma'-1}\theta^{2-\gamma'}\bs{\hat{\theta}}.
\end{equation}
It is straightforward to check, by plugging this result into the lens equation, $ \bs{\beta} = \bs{\theta} - \bs{\alpha}(\bs{\theta}) $, and solving for $ \beta=0 $, that $ \theta_{E} $ is indeed the Einstein radius.
The lens equation can be written in a dimensionless form, in $ \theta_{E} $ units, as
\begin{equation}
\bs{y} = \bs{x} - |x|^{2-\gamma'}\frac{\bs{x}	}{|x|},
\label{eq:lenseq}
\end{equation}
where $ \bs{x}\equiv\bs{\theta}/\theta_{E} $ and $ \bs{y}\equiv\bs{\beta}/\theta_{E} $ are the dimensionless angular position and impact parameter, respectively. Thus, Eq.~\eqref{eq:lenseq} relates the observed angular positions of the images to the lens parameter $ \theta_{E} $ and $ \gamma' $ and the source angular position, or---for the simulations carried out in this work---given the lens parameters and the source position, the images positions can be determined.

\end{document}